\documentclass[aps,prc,showpacs,floatfix,preprint]{revtex4-1}
\usepackage{graphicx}
\usepackage{epsfig}
\usepackage{color}
\usepackage[normalem]{ulem}  

\begin{document}
 \title{Low density nuclear matter with quantum molecular dynamics:  The role of the symmetry energy}
 \author{Rana Nandi}
 \email[]{nandi@fias.uni-frankfurt.de}
 \affiliation{Frankfurt Institute for Advanced Studies, 60438 Frankfurt am Main, Germany}
 
 \author{Stefan Schramm}
  \email[]{schramm@fias.uni-frankfurt.de}
  \affiliation{Frankfurt Institute for Advanced Studies, 60438 Frankfurt am Main, Germany}
  
  \begin{abstract}
   We study the effect of isospin-dependent nuclear forces on the pasta phase in the inner crust of neutron stars.
   To this end we model the crust within the framework of quantum molecular dynamics (QMD). For maximizing
   the numerical performance, a newly developed code has been implemented on GPU processors.
   As a first application of the crust studies we investigate the dependence of the particular pasta phases on 
   the isospin dependence of the interaction, including non-linear terms in this sector of the interactions.
   Our results indicate that in contrast to earlier studies the phase diagram of the pasta phase is not very
   sensitive to isospin effects. We show that the extraction of the isospin parameters like asymmetry energy
   and slope from numerical data is affected by higher-order terms in the asymmetry dependence of the energies per particle.
   Furthermore, a rapid transition from the pasta to a homogeneous phase is observed even for proton-to-neutron ratios typical for a supernova environment.

   \end{abstract}

  \maketitle

\section{Introduction}
In the crust of neutron stars, at densities well below nuclear saturation density nuclei form 
crystalline structures embedded in an electron plasma in order to minimize the Coulomb energy. 
At higher densities, when nuclei are about to dissolve into uniform matter, 
various interesting spatial structures such as cylindrical and slab shaped nuclei and cylindrical and 
spherical bubbles etc., collectively called nuclear ``pasta'', may appear \cite{Ravenhall83,Hashimoto84}. 

The study of the pasta phase is very important for various astrophysical reasons. In core-collapse supernovae the  pasta phase
significantly affects neutrino transport through the matter, which plays a crucial role in the eventual supernova
explosion \cite{Horowitz04}. On the other hand electron-pasta scattering has a strong impact on the transport 
properties like electrical and thermal conductivities of the neutron star crustal matter. An enhanced electrical
resistivity due to the pasta structures could be a central effect to explain the decay of magnetic field in
neutron stars \cite{Pons13} and the thermal conductivity is essential to understand the cooling behaviour of 
these stars \cite{Newton13, Horowitz15}. The presence of the pasta phase might also be important to understand
the mechanism of pulsar glitches \cite{Lorenz93, Watanabe07}.

So far, a number of authors have studied the properties of the pasta phase. Most of the studies adopt static
methods such as liquid-drop models \cite{Lorenz93, Watanabe00}, Thomas-Fermi approximations
\cite{Oyamatsu93, Lassaut87} and the Hartee-Fock method \cite{Gogelein07, Newton09}. In these models few specific
shapes are assumed and free energies are calculated for all the shapes as a function of baryon density.
The  equilibrium shape at a particular density is then determined by minimizing the free energy. 
However, to study the formation and evolution of the pasta phase one needs to employ a dynamical approach that
allows for arbitrary nuclear shapes and can incorporate the thermal fluctuations on the nucleon distribution in 
a natural way. Furthermore, as the nuclear matter is a typical frustrated system with competing attractive
nuclear and repulsive Coulomb forces, many energetically competing structures might occur.
Only a few groups so far have adopted a dynamical approach.
The first study in this direction was done by Maruyama {\it et al} \cite{Maruyama98}, who developed a quantum
molecular dynamics (QMD) model to study the pasta phase. Later Watanabe {\it et al} adopted this QMD model and 
studied characteristics of the pasta at zero temperature \cite{Watanabe03} as well as finite temperatures
\cite{Watanabe04} and also the transition between different nuclear shapes \cite{Watanabe05, Watanabe09}.
Horowitz {\it et al} developed a semi-classical dynamic model (SMD) and studied various transport properties
\cite{Horowitz04,Horowitz05, Horowitz08, Horowitz09, Chugonov10, Horowitz15} as well as formation
\cite{Schneider13} and characteristics \cite{Schneider14} of the pasta phase. Recently, Dorso {\it et al}
\cite{Dorso12, Molinelli14} and Sch\"utrumpf {\it et al} \cite{Schuetrumpf13} studied the pasta phase using 
classical molecular dynamics (CMD) and a time-dependent Hartree-Fock approach, respectively.

The nuclear symmetry energy and its density dependence play crucial roles in both nuclear physics and astrophysics
\cite{Lattimer07,Li08}. Laboratory experiments constrain the symmetry energy at saturation density quite well
around $30\pm4$ MeV, but its slope $L$ at saturation is still very uncertain and is expected to lie in the range $20-120$ MeV
\cite{Zhang13}. The effect of different $L$ on the pasta phase of inner crust matter of neutron stars has
been studied within the liquid drop model
\cite{Bao14}  as well as the Thomas-Fermi approximation \cite{Oyamatsu07, Grill12, Bao15}. In all of these
calculations it was found that the width of the pasta phase decreases with increasing value of $L$. Sonoda
{\it et al} \cite{Sonoda08} used two different QMD models with different $L$ to study the pasta phase with proton
fraction $Y_p=0.3$ and found behaviour similar to the static calculations. But to understand the dependence of
the pasta phase on the asymmetry dependence of the matter alone, one has to do such a study consistently within the same nuclear model approach. 
Therefore, in this article we study the dependence of the pasta phase on the asymmetry properties within a single QMD
model. The article is structured in the following way. After outlining the general formalism in Sec. II, we
present a careful parameter study of different strengths of the isospin forces in Sec. III. Finally, in Sec. IV we
draw conclusions and present an outlook of upcoming work in this direction.

\section{Formalism}
In QMD the state of a nucleon is represented by a Gaussian wave packet given as (we set $\hbar=c=1$)
\begin{equation}
 \psi({\bf r_i}) = \frac{1}{(2\pi C_W)^{3/4}} \exp\left[-\frac{({\bf r_i - R_i})^2}{4C_W}+ i \,{\bf r\cdot P_i}\right],
\end{equation}
where ${\bf R}_i$ and ${\bf P}_i$ denote the center of the position and momentum of the wave packet $i$, respectively, with the corresponding width $C_W$. Then the total wave function for the $N$-nucleon system is obtained by taking the direct product of single-nucleon wave functions
\begin{equation}
 \Psi(\{{\bf r}\}) =  \prod_i^N \psi({\bf r_i}) 
\end{equation}

Here we adopt an effective interaction developed by Maruyama {\it et al.} \cite{Maruyama98}, to simulate
the nuclear matter at sub-saturation densities. The Hamiltonian of the interaction has several terms:
\begin{equation}
 {\cal H}=T+V_{\rm Pauli}+V_{\rm Skyrme}+V_{\rm sym}+V_{\rm MD}+V_{\rm Coul},
\end{equation}
where $T$ is the kinetic energy and $V_{\rm Pauli}$ is the phenomenological Pauli potential which effectively mimics the Pauli exclusion
principle. $V_{\rm Skyrme}$ is the nucleon-nucleon potential similar to Skyrme-like interactions, $V_{\rm sym}$ is the isospin-dependent potential related to the symmetry
energy, and $V_{\rm MD}$ represents the momentum-dependent potential that arise from the Fock terms of Yukawa-type interactions.
Finally, $V_{\rm Coul}$ is the Coulomb
potential. The explicit expressions for all the terms are as follows
\begin{widetext}
\begin{eqnarray}
  T  &=& \sum_{i, j(\ne i)} \frac{\bf P_{\it i}^{2}}{2 m_{i}}\ ,\label{kin}\\  
  V_{\rm Pauli} &=& 
  \frac{C_{\rm P}}{2}\
  \left( \frac{1}{q_0 p_0}\right)^3
  \sum_{i, j(\neq i)} 
  \exp{ \left [ -\frac{({\bf R}_i-{\bf R}_j)^2}{2q_0^2} 
          -\frac{({\bf P}_i-{\bf P}_j)^2}{2p_0^2} \right ] }\
  \delta_{\tau_i \tau_j} \delta_{\sigma_i \sigma_j}\ ,\label{pauli}\\
  V_{\rm Skyrme} &=&
  {\alpha\over 2\rho_0}\sum_{i, j (\neq i)}
  \rho_{ij}
  +  {\beta\over (1+\tau)\ \rho_0^{\tau}}
  \sum_i \left[ \sum_{j (\neq i)} 
                     \tilde{\rho}_{ij}  \right]^{\tau}\ ,
                   \label{skyrme}\\
   V_{\rm sym} &=&
  {C_{\rm s}^{(1)}\over 2\rho_0} \sum_{i , j(\neq i)} \,
  ( 1 - 2 | \tau_i - \tau_j | ) \ \rho_{ij} 
   + {C_{\rm s}^{(2)}\over (1+\gamma)\rho_0^\gamma} \sum_i \left[\sum_{j(\neq i)} \,
  ( 1 - 2 | \tau_i - \tau_j | ) \ \bar{\rho}_{ij}\right]^\gamma  \ ,\label{sym}\\
  V_{\rm MD}  &=&
         {C_{\rm ex}^{(1)} \over 2\rho_0} \sum_{i , j(\neq i)} 
      {1 \over 1+\left[{{\bf P}_i-{\bf P}_j \over \mu_1}\right]^2} 
      \ \rho_{ij}
     +   {C_{\rm ex}^{(2)} \over 2\rho_0} \sum_{i , j(\neq i)} 
      {1 \over 1+\left[{{\bf P}_i-{\bf P}_j \over  \mu_2}\right]^2} 
      \ \rho_{ij}\ ,\label{md}\\
  V_{\rm Coul} &=&
  {e^2 \over 2}\sum_{i , j(\neq i)}
  \left(\tau_{i}+\frac{1}{2}\right) \, \left(\tau_{j}+\frac{1}{2}\right)
  \int\!\!\!\!\int d^3{\bf r}\,d^3{\bf r}^{\prime} 
  { 1 \over|{\bf r}-{\bf r}^{\prime}|} \,
  \rho_i({\bf r})\rho_j({\bf r}^{\prime})\ ,\label{coulomb}
\end{eqnarray}
\end{widetext}
where $\sigma_{i}$ and $\tau_{i}$ ($1/2$ for protons and $-1/2$ for neutrons) are the nucleon spin and isospin, respectively and
$\rho_{ij}$, $\tilde{\rho}_{ij}$ and $\bar{\rho}_{ij}$  represent the overlap between single-nucleon densities and defined as
\begin{equation}
  \rho_{ij} \equiv \int { d^3{\bf r} \rho_i({\bf r})
  \rho_j({\bf r}) }\ ,\quad \tilde{\rho}_{ij} \equiv \int { d^3{\bf r} \tilde{\rho_i}({\bf r})\tilde{ \rho_j}({\bf  r})}\ ,
  \quad \bar{\rho}_{ij} \equiv \int { d^3{\bf r} \bar{\rho_i}({\bf r})\bar{ \rho_j}({\bf  r})},\
  \label{rhoij}
\end{equation}
whereas the single-nucleon densities are given by 
\begin{eqnarray}
  \rho_i({\bf r}) & = & \left| \psi_{i}({\bf r}) \right|^{2}
  = \frac{1}{(2\pi C_W)^{3/2}}\ \exp{\left[
                - \frac{({\bf r} - {\bf R}_i)^2}{2C_W} \right]}\ ,\quad \\
  \tilde{\rho_i}({\bf r}) & = &
  \frac{1}{(2\pi \tilde{C}_W)^{3/2}}\ \exp{\left[
                - \frac{({\bf r} - {\bf R}_i)^2}{2\tilde{C}_W} \right]}\ , \\
   \bar{\rho_i}({\bf r}) & = &
  \frac{1}{(2\pi \bar{C}_W)^{3/2}}\ \exp{\left[
                - \frac{({\bf r} - {\bf R}_i)^2}{2\bar{C}_W} \right]}\ ,             
\end{eqnarray}
with
\begin{equation}
  \tilde{C}_W = \frac{1}{2}(1+\tau)^{1/ \tau}\ C_W \quad {\rm and}\quad \bar{C}_W = \frac{1}{2}(1+\gamma)^{1/ \gamma}\ C_W \ .
\end{equation}
The modified widths $\tilde{C}_W$ and $\bar{C}_W$ of the Gaussian
wave packet are introduced to adjust the effect of density-dependent terms (for more details see Ref. \cite{Maruyama98}).
In the original model of Maruyama {\it et al.} \cite{Maruyama98} there was only the linear term in $V_{\rm sym}$. To study the 
density dependence of the symmetry energy we have added a second non-linear term analogously to the density-dependent term appearing
in the isospin-0 Skyrme potential $V_{\rm Skyrme}$. Out of the 13 parameters (Table \ref{parameter} and \ref{symparams}) of the model 10 are obtained from the properties of nuclear matter at saturation. The Gaussian width $C_W$ is chosen to get a good fit to the binding 
energies of finite nuclei. The symmetry energy coefficients $C_{S}^{(1)}$ and $C_{S}^{(2)}$ are free parameters and are adjusted to
achieve reasonable values of the symmetry energy and its slope ($L$) at saturation (see later discussion).

\begin{table}[]
\caption{Parameter set for the interaction \cite{Maruyama98}}
{\small \begin{tabular}{cccc}
\hline\hline 
& $C_{\rm P}$ (MeV) &\qquad\qquad 207 &\\
& $p_{0}$ (MeV/$c$) &\qquad\qquad 120 &\\
& $q_{0}$ (fm) &\qquad\qquad 1.644 &\\
& $\alpha$ (MeV) &\qquad\qquad $-92.86$ &\\
& $\beta$ (MeV) &\qquad\qquad 169.28 &\\
& $\tau$ &\qquad\qquad 1.33333 &\\
& $C_{\rm ex}^{(1)}$ (MeV) &\qquad\qquad $-258.54$ &\\
& $C_{\rm ex}^{(2)}$ (MeV) &\qquad\qquad 375.6 &\\
& $\mu_1$ (fm$^{-1}$) &\qquad\qquad 2.35 &\\
& $\mu_2$ (fm$^{-1}$) &\qquad\qquad 0.4 &\\
& $C_W$ (fm$^2$) &\qquad\qquad 2.1 &\\
\hline\hline 
\end{tabular}}
\label{parameter}
\end{table}
\begin{table}[]
\caption{Symmetry energy coefficients}
\begin{tabular}{ccccccc}
   Set\qquad\quad &$C_s^{(1)}$(MeV)\qquad\qquad & $C_s^{(2)}$(MeV)\qquad\qquad& $\gamma$\qquad\qquad & $e_{\rm sym}(\rho_0)(\rm MeV)$\qquad\qquad & $L\,(\rm MeV)$ \\ \hline\hline 
   I\qquad\qquad&$30.0$  \qquad\qquad   & $-15.0 $\qquad\qquad   &  $3.0$\qquad\qquad   &$34.6$\qquad\qquad         &$76.8$\\
   II\qquad\qquad&$25.0$  \qquad\qquad   & $0.0$   \qquad\qquad   &  $0.0$\qquad\qquad   &$34.3$\qquad\qquad         &$91.8$\\
   III\qquad\qquad&$18.0$  \qquad\qquad   & $22.5$  \qquad\qquad   &  $3.0$\qquad\qquad   &$34.2$\qquad \qquad&$114.4$\\   
 \end{tabular}
 \label{symparams}
\end{table}

In order to obtain the equilibrium configuration we use following equations of motion with damping terms \cite{Maruyama98}:
\begin{eqnarray}
 {\bf\dot{R}_i } &=& \frac{\partial H}{\partial {\bf P_i}} - \mu_R\frac{\partial H}{\partial {\bf R_i}},\nonumber\\ 
 {\bf\dot{P}_i } &=& -\frac{\partial H}{\partial {\bf R_i}} - \mu_P\frac{\partial H}{\partial {\bf P_i}},\label{eom}
\end{eqnarray}
where $\mu_R$ and $\mu_P$ are damping coefficients, which are positive definite and relate to the relaxation time scale. 

\section{Results}
\subsection{Simulation procedure}
Adopting the theoretical framework outlined in Sec. II
we have carried out QMD simulation of a system containing neutrons, protons and electrons at essentially zero temperature.
The particles are confined in a cubic box, the size of which is determined from a given particle number ($ {\cal N} $) and the average 
density ($\rho_{\rm av}$). 
To simulate infinite nuclear matter we impose periodic boundary conditions. We include  2048 nucleons, out of which 608 are
protons and 1440 are neutrons, such that the proton fraction ($Y_p$) is close to 0.3, a value relevant for studies of
core-collapse supernovae. We also simulate symmetric nuclear matter with an equal number of protons and neutrons ($Y_p=0.5$).
The number of protons (neutrons) with spin-up are taken to be equal to that of protons (neutrons) with spin-down. To calculate
the Coulomb interaction we employ the Ewald method \cite{Watanabe03}, where electrons are considered to form  a uniform background and
make the system charge neutral.

As an initial configuration we distribute nucleons randomly in phase space. Then with the help of the Nos\'{e}-Hoover 
thermostat \cite{Watanabe04} we equilibrate the system at $T=20$ MeV for about $2000$ fm/c. To achieve the ground state
configuration we then slowly cool down the system in accordance with the damped equations of motion (Eqs. \ref{eom}) until the 
temperature reaches a value below $1$ keV. 

For speeding up the simulation we ported the QMD code to a GPU version, making full use of the nearly 3600 cores in the AMD FirePro
S10000 graphics processor unit. With this implementation we can reach the ground state, which requires $\sim 10^4$ fm/c, within  
a few hrs of computational time. All the simulations are done at the LOEWE-CSC CPU/GPU cluster at Frankfurt University.

\subsection{Finite nuclei and  asymmetric nuclear matter}
\begin{figure}
 \centering
 \includegraphics[width=0.6\textwidth]{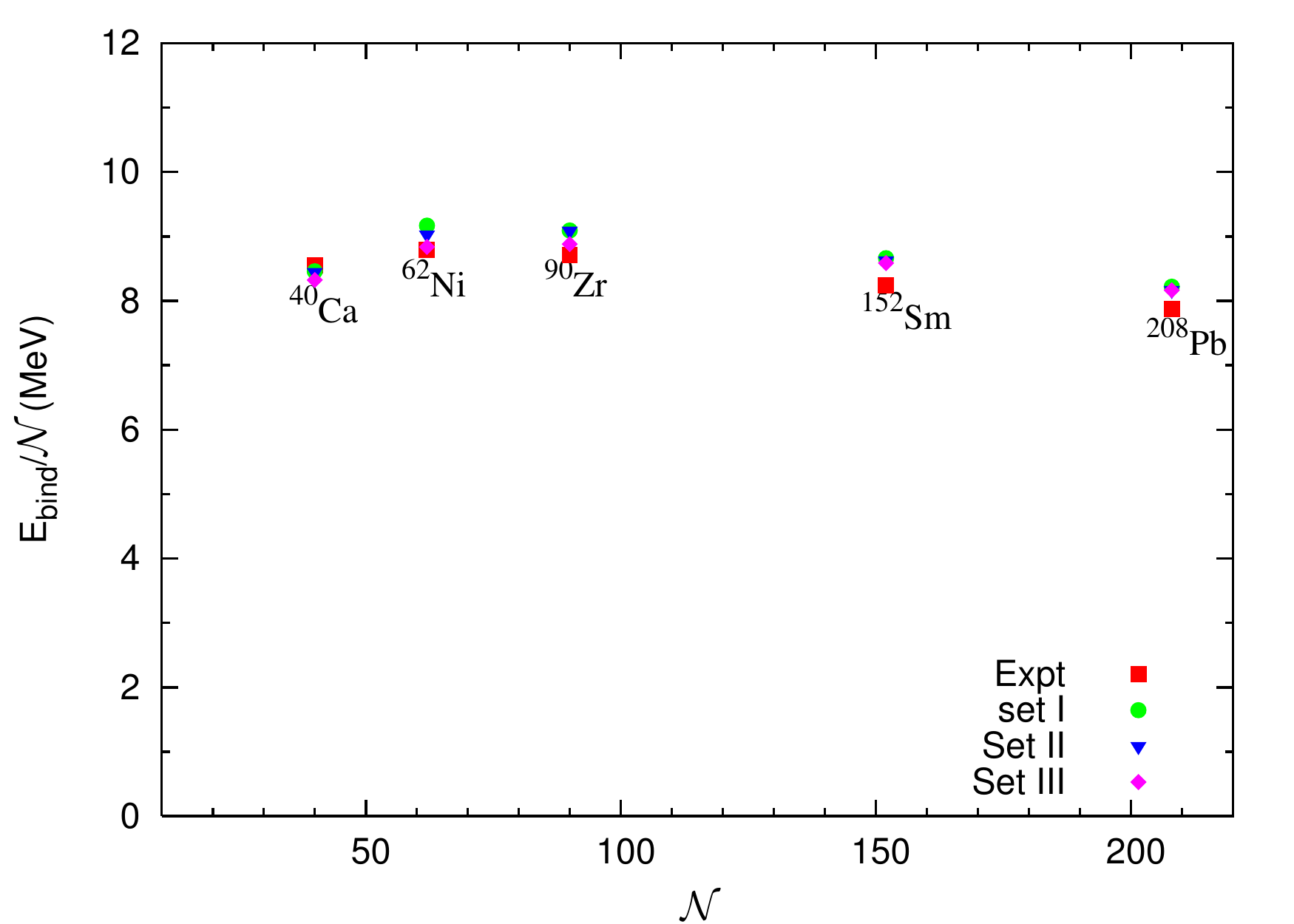}
 \caption{Binding energies for nuclei obtained from simulation for three different parameter sets: (I) $C_s^{(1)}=30.0,\, C_s^{(2)}=-15.0$;
          (II) $C_s^{(1)}=25.0,\, C_s^{(2)}=0.0$; (III) $C_s^{(1)}=18.0,\, C_s^{(2)}=22.5$. Experimental values are denoted by solid squares.}
 \label{f:be}
\end{figure}

We have chosen three different parameter sets corresponding to three different sets of values  for the coefficients $C_S^{(1)}$ and 
$C_S^{(2)}$ (see Table \ref{symparams}). 
In order to check the validity of our parameter sets, in Fig. \ref{f:be}, we show the binding energies of the ground state of a number of 
nuclei, covering a range of nuclear masses from Ca to Pb, obtained from our simulation using all three parameter sets.
Given the realistically achievable accuracy within a molecular dynamics approach, all of them match the experimental values reasonably well.

\begin{figure}
  \begin{center}
    \begin{tabular}{ccc}
      \resizebox{45mm}{!}{\includegraphics{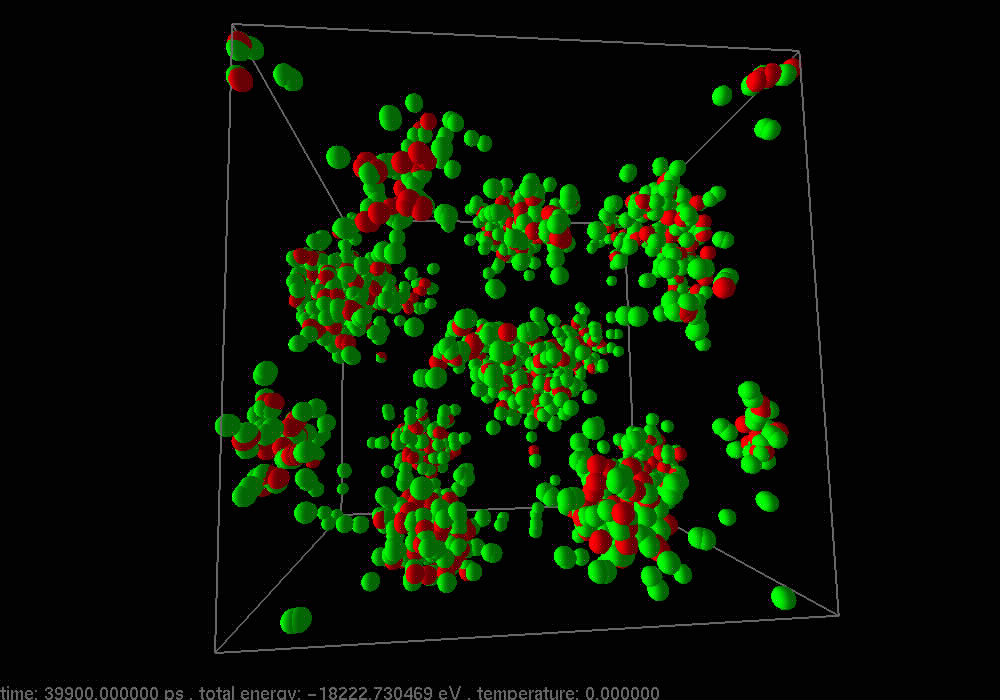}}&
      \resizebox{45mm}{!}{\includegraphics{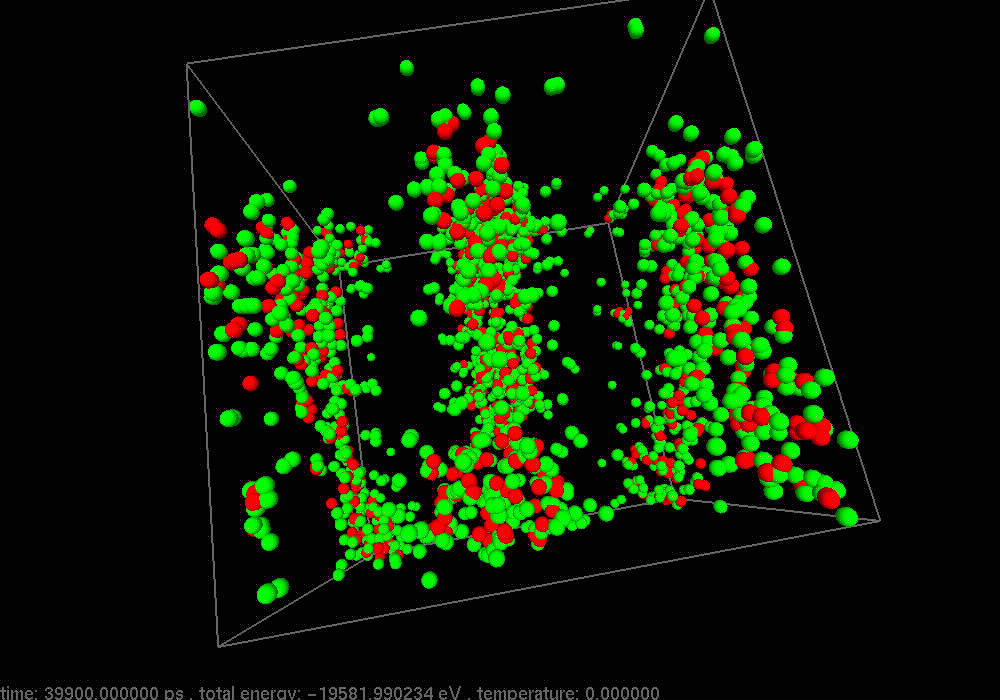}}&
      \resizebox{45mm}{!}{\includegraphics{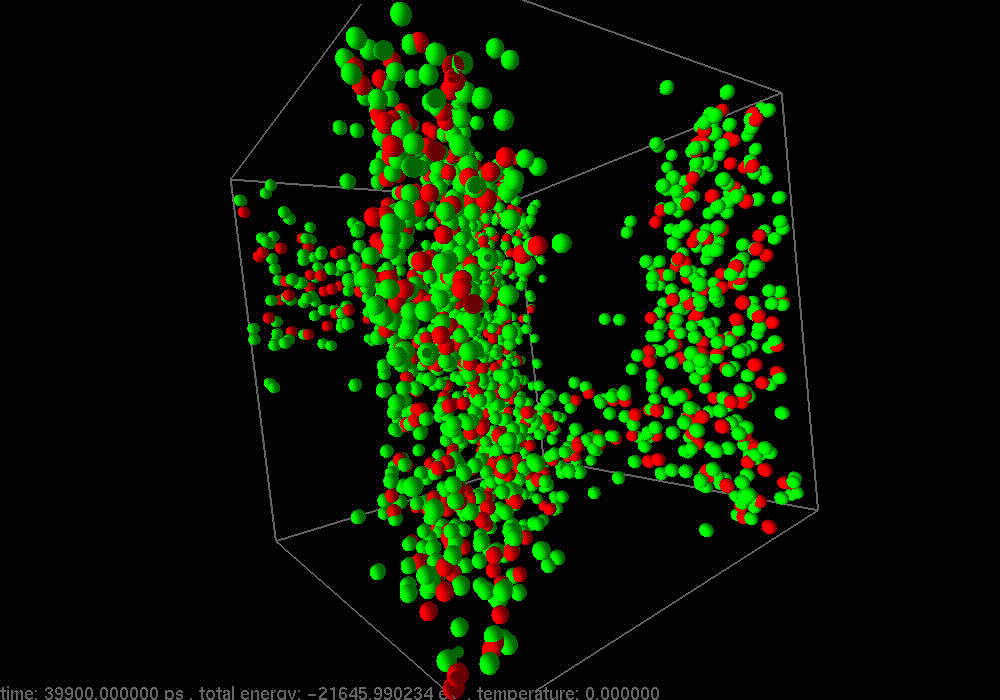}} 
    \end{tabular}
  \end{center}
  \begin{center}
    \begin{tabular}{cc}  
      \resizebox{45mm}{!}{\includegraphics{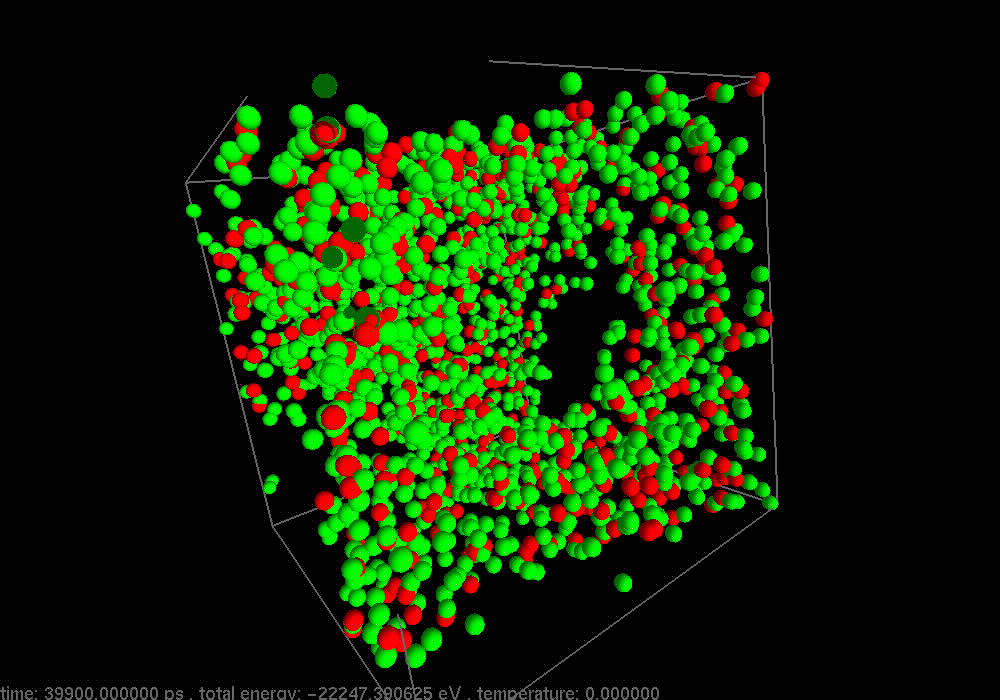}} &
      \resizebox{45mm}{!}{\includegraphics{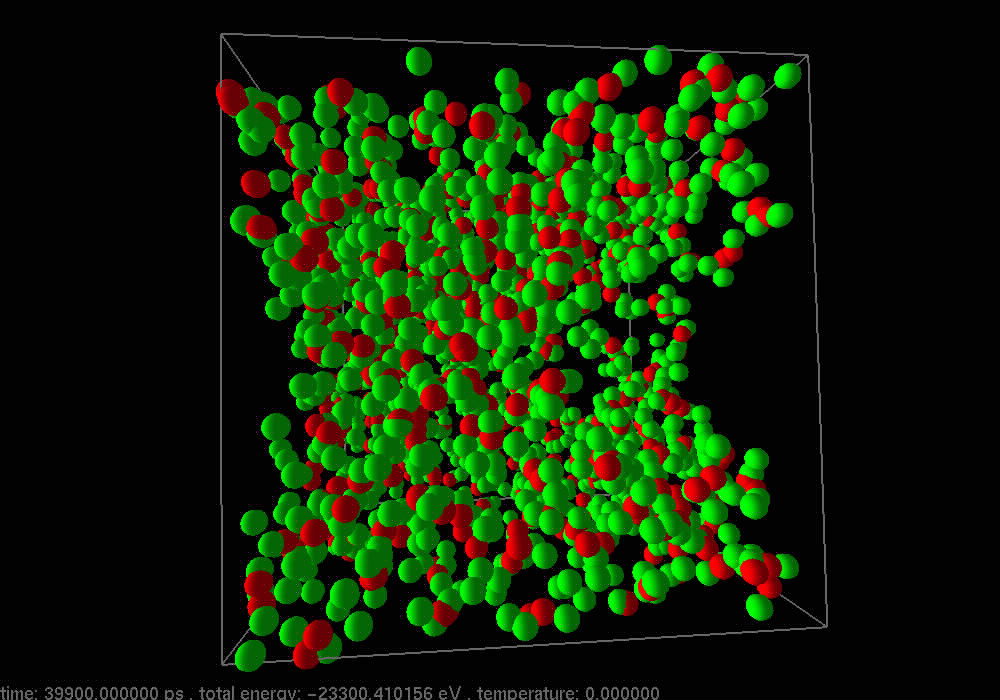}}
    \end{tabular}
    \caption{The nucleon distribution of phases with simple structures of cold matter at $Y_p=0.3$ and densities (left to right)
             $0.1\,\rho_0$, $0.2\,\rho_0$, $0.36\,\rho_0$, $0.5\,\rho_0$ and $0.575\,\rho_0$, respectively. Green (red) spheres represent
             neutrons (protons).}
    \label{snapshots}
  \end{center}
\end{figure}

Turning to increasingly asymmetric matter, in Fig. \ref{snapshots} the snapshots for the nucleon distributions of various phases 
for $Y_p=0.3$ are shown for the parameter set II. It is observed that all the regular pasta shapes i.e. sphere, cylinder, slab, cylindrical hole, 
spherical hole with increasing density, are reproduced successfully as in the earlier investigation \cite{Watanabe03}. Similar results
are also obtained for other two parameter sets.

\begin{figure}
 \centering
 \includegraphics[width=0.6\textwidth]{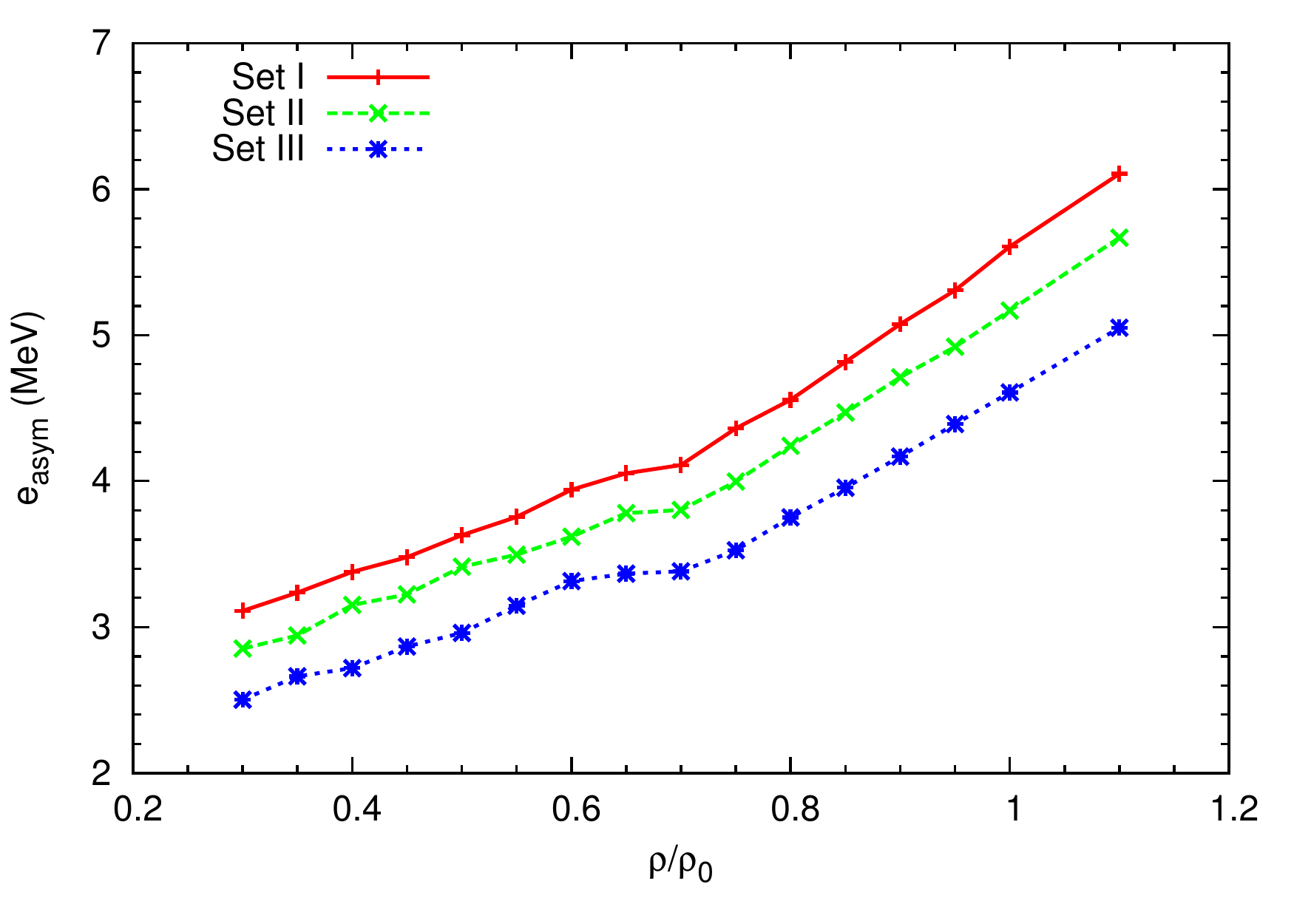}
 \caption{Asymmetry energy per nucleon obtained from simulation as a function of density for three different parameter sets.}
 \label{f:asym}
\end{figure}
In Fig. \ref{f:asym} we plot the asymmetry energy per nucleon which we define as:
\begin{equation}
 e_{\rm asym}(\rho) = e(\rho, Y_p=0.3) - e(\rho,Y_p=0.0), \label{asym}
\end{equation}
where $e(\rho, Y_p=0.0)$ is the energy per nucleon for symmetric matter and $e(\rho, Y_p=0.3)$ is that of matter with $Y_p=0.3$.
The figure shows that three different parameter sets lead to distinctly different asymmetry energies at all densities. 

\subsection{Determination of $e_{{\rm sym}(\rho_0)}$ and $L$ (parabolic approximation)}
Fig. \ref{f:asym} is a direct result from our simulation. To make connection with other analyses of isospin effects we next derive 
commonly used quantities such as the symmetry energy $e_{\rm sym}(\rho_0)$ and its slope $L$ at saturation density. 
The energy per nucleon of asymmetric nuclear matter can be written as a Taylor series with respect to the neutron excess
$\delta = (\rho_n - \rho_p)/(\rho_n + \rho_p)$, where $\rho_n$ and $\rho_p$ are the neutron and proton
densities, respectively. A commonly used approach retains only the lowest-order non-vanishing term in $\delta$ (parabolic approximation):
\begin{equation}
 e(\rho,\delta) = e_0(\rho) + e_{\rm sym}(\rho)\delta^2, \label{epar2}
\end{equation}
where $e_0(\rho) = e(\rho,\delta=0)$ is the energy per nucleon of symmetric matter and $e_{\rm sym}(\rho)$ is the
nuclear symmetry energy. The symmetry energy can then be expanded (to lowest order) around the normal nuclear density {$\rho_0$} as
\begin{equation}
 e_{\rm sym}(\rho) = e_{\rm sym}(\rho_0) + L\, \chi\, ,\label{esym2}
\end{equation}
where $\chi = (\rho-\rho_0)/3\rho_0$ denotes the deviation from $\rho_0$ and $L$ is the slope of the symmetry energy at {$\rho_0$} given by
\begin{equation}
 L = 3\rho_0\frac{\partial e_{\rm sym}(\rho)}{\partial \rho}\Bigg|_{\rho=\rho_0}\label{slope}\, .
\end{equation}

\begin{figure}
  \begin{center}
    \begin{tabular}{ccc}
      \resizebox{52mm}{!}{\includegraphics{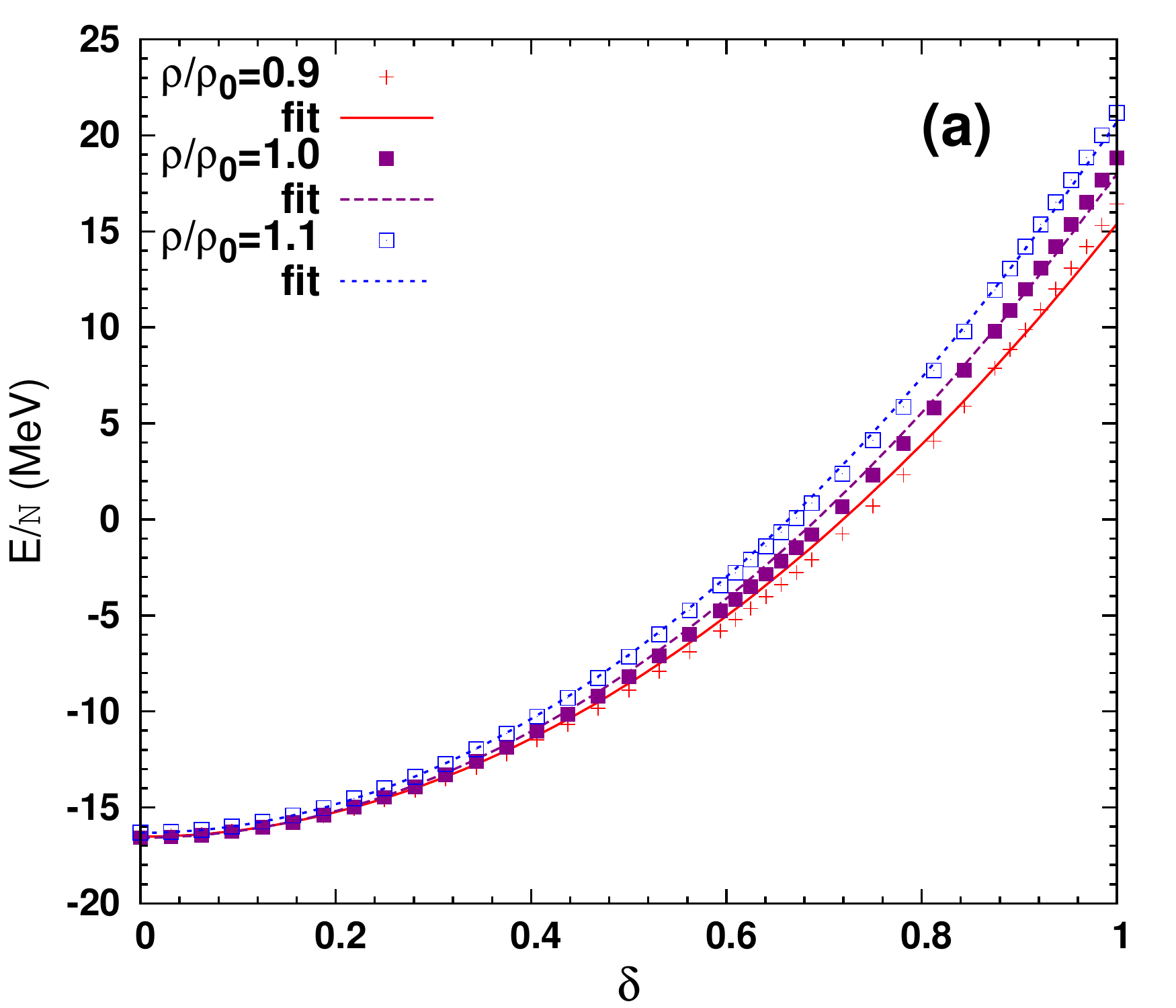}} &
      \resizebox{52mm}{!}{\includegraphics{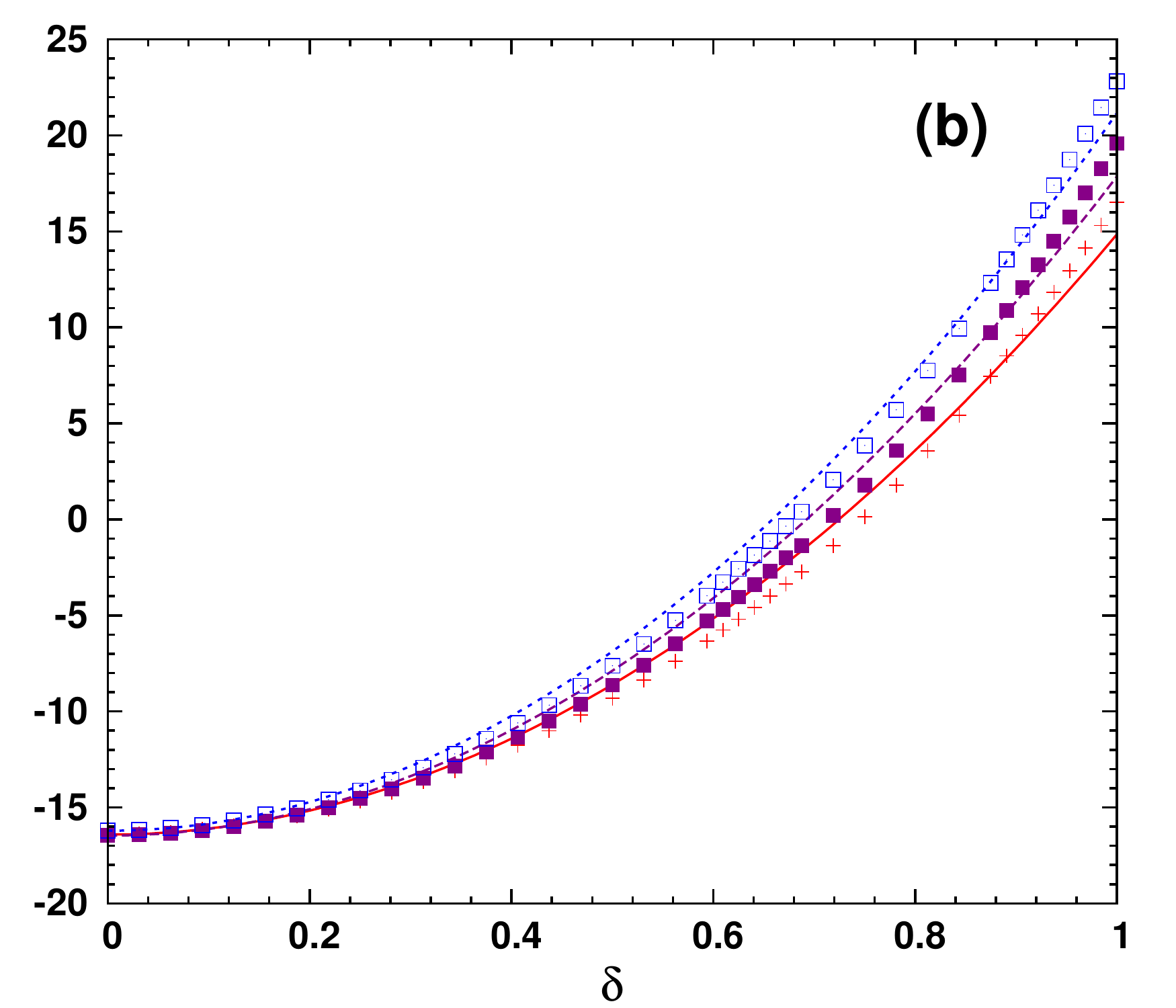}}&
      \resizebox{52mm}{!}{\includegraphics{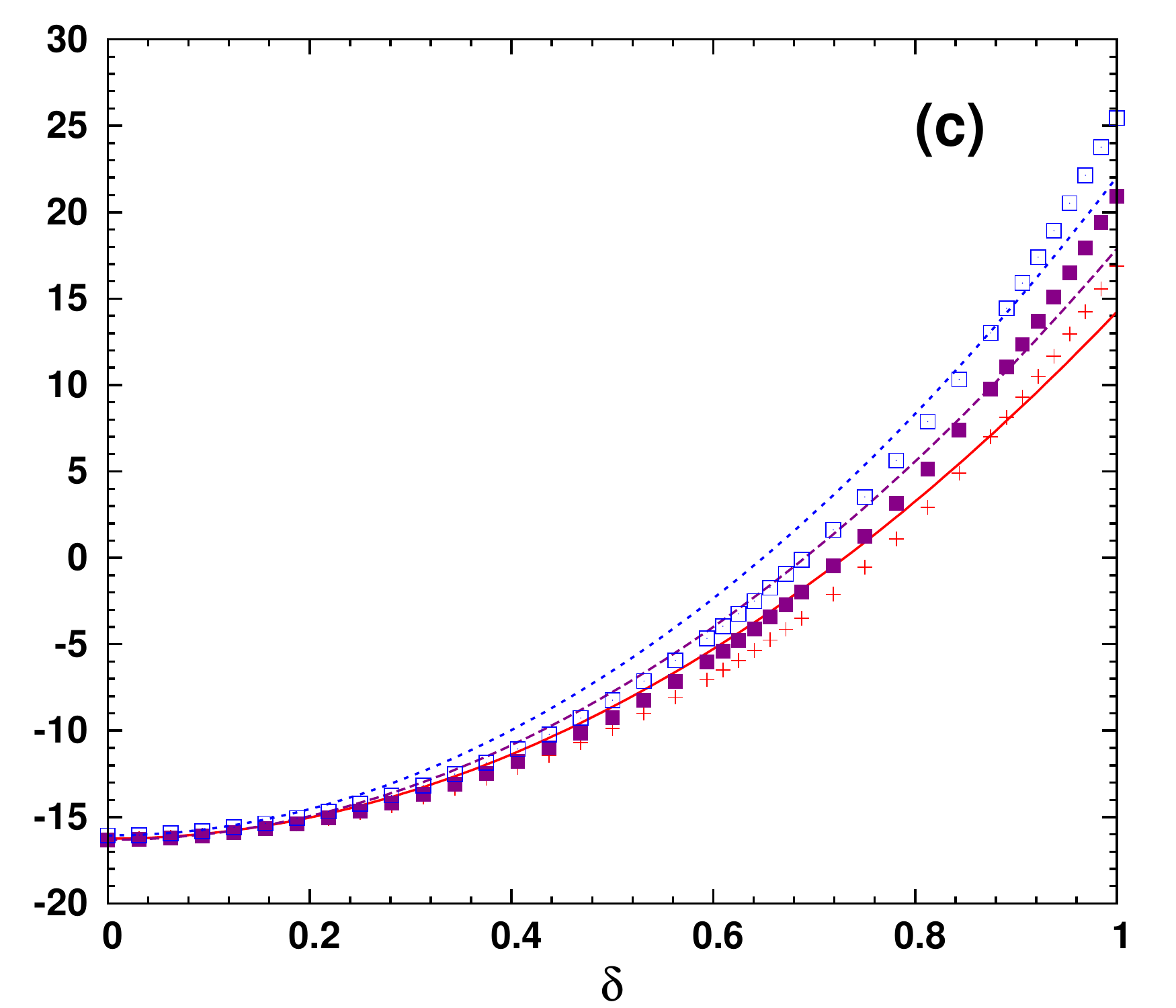}} 
    \end{tabular}
    \caption{Fit of energy per nucleon using Eq. (\ref{epar2}) for different parameter sets: (a) $C_s^{(1)}=30.0,\, C_s^{(2)}=-15.0$;
            (b) $C_s^{(1)}=25.0,\, C_s^{(2)}=0.0$; (c) $C_s^{(1)}=18.0,\, C_s^{(2)}=22.5$.}
     \label{f:fit1}
  \end{center}
\end{figure}

To evaluate $e_{\rm sym}(\rho_0)$ and $L$ we run our simulation for different values of $\delta$ (from 0 to 1) keeping the density 
($\rho_{\rm av}$) fixed at $\rho_0$ and switching off the Coulomb interaction, for all three sets of values of $C_s^{(1)}$ and $C_s^{(2)}$
given in Table \ref{symparams}.
Then we fit the obtained values of energy per nucleon with Eq. (\ref{epar2}) and obtain $e_{\rm sym}(\rho_0)$ as fit 
parameter (see Fig. \ref{f:fit1}). Following the same procedure we also determine $e_{\rm sym}(0.9\rho_0)$ and $e_{\rm sym}(1.1\rho_0)$,
which are then used to calculate $L$ as
\begin{equation}
  L =  3\rho_0\frac{e_{\rm sym}(1.1\rho_0)-e_{\rm sym}(0.9\rho_0)}{1.1\rho_0-0.9\rho_0}\,. \label{slope_sim}
\end{equation}
In Fig. \ref{f:fit1} we plot the energy per nucleon obtained from the simulation as well as from the fitting procedure, as a function of $\delta$ for 
three different densities and for three different sets of parameters $C_s^{(1)}$ and $C_s^{(2)}$. The resulting values of $e_{\rm sym}(\rho_0)$
and $L$ are given in Table \ref{symparams}. From the table one can infer that, although the symmetry energies at saturation density
are not very different, we get three different values for its slope namely 77, 92 and 114 MeV, respectively. 

\subsection{Minkowski functionals}
To quantify various nuclear shapes obtained from the simulations we calculate the corresponding Minkowski functionals \cite{Michielsen01}. In three 
dimensions any arbitrary shape can be characterised by four Minkowski functionals: volume $V$, surface area $A$, integral
mean curvature $H$ and Euler characteristic $\chi$. The last two quantities are determined from the principal curvatures $\kappa_1$ and $\kappa_2$
on the surface $\partial K$ as
\begin{equation}
 H = {1\over 2}\int_{\partial K}(\kappa_1 + \kappa_2)dA \qquad \chi = \frac{1}{2\pi}\int_{\partial K}\kappa_1\cdot\kappa_2\,dA \, .\label{mf}
\end{equation}
The Euler characteristic can also be calculated from the topology of the structure as \cite{Michielsen01}
\begin{equation}
 \chi = {\rm number\ of\ connected\ regions} + {\rm number\ of\ cavities} - {\rm number\ of\ tunnels} \,.\label{ec}
\end{equation}

To calculate the Minkowski functionals we first divide the simulation box in $64^3$ voxels and calculate densities at each voxel ($j$) as 
$\rho_j^{n,p} = \sum_i^N\rho_i^{n,p} $. We choose a density threshold ($\rho_{\rm th}$) and turn the density field into a
black-and-white data set according to
\begin{eqnarray}
 {\rm voxel}\ j &=& {\rm black\ if}\ \rho_j \geq \rho_{\rm th} \nonumber \\
        &=& {\rm white\ if}\ \rho_j < \rho_{\rm th}\,.
\end{eqnarray}
Then we apply the marching cube algorithm \cite{Lorensen87} to create a smooth polygonal surface representation of the black voxels.
Finally, the Minkowski functionals for the polygon are evaluated with the help of the Karambola package \cite{karambola}.
In order to investigate the dependence of the results on the choice of threshold, we repeat the procedure for a range of densities 
$\rho_{\rm th}$. 
\begin{figure}
 \centering
 \includegraphics[width=0.6\textwidth,angle=-90]{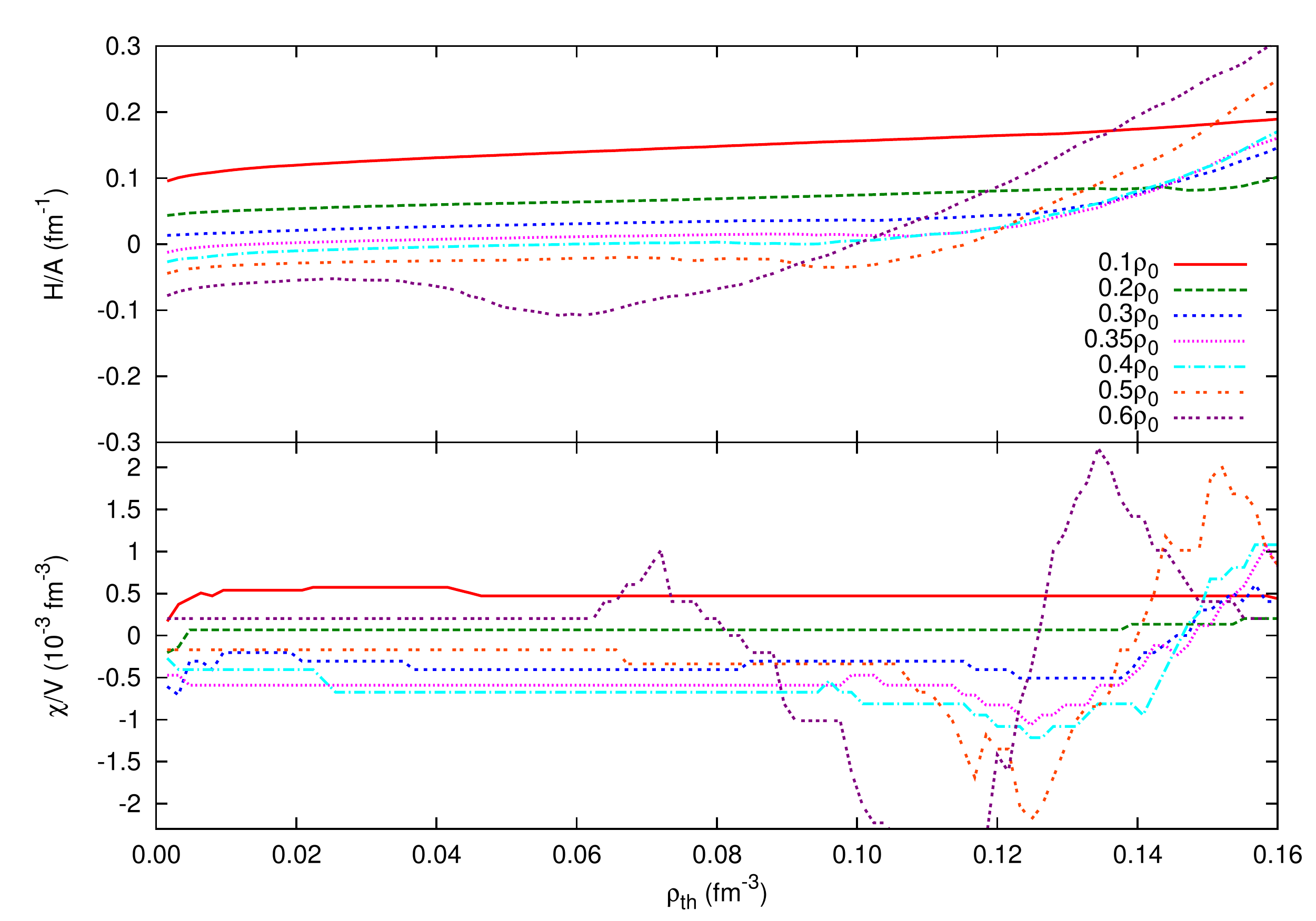}
 \caption{Normalised mean curvature and Euler characteristic as functions of threshold density for various nucleon
 densities}
 \label{f:mf_th}
\end{figure}
In Fig. \ref{f:mf_th} we show
normalised Minkowski functionals $H/A$ and $\chi/V$ for various nucleon densities as functions of the threshold density. 
One can observe that $H/A$ increases slowly with $\rho_{\rm th}$, while the slope decreases with increased density. For $\chi/V$, we see
that a plateau region exists ($\sim 0.02-0.08$ fm$^{-3}$) that covers all reasonable values of $\rho_{\rm th}$. 
The width of this plateau decreases with density. For our analysis we take 
the average value across the plateau and evaluate the corresponding standard deviation for obtaining an error estimate.

\begin{figure}
 \begin{center}
  \begin{tabular}{cc}
   \resizebox{70mm}{!}{\includegraphics{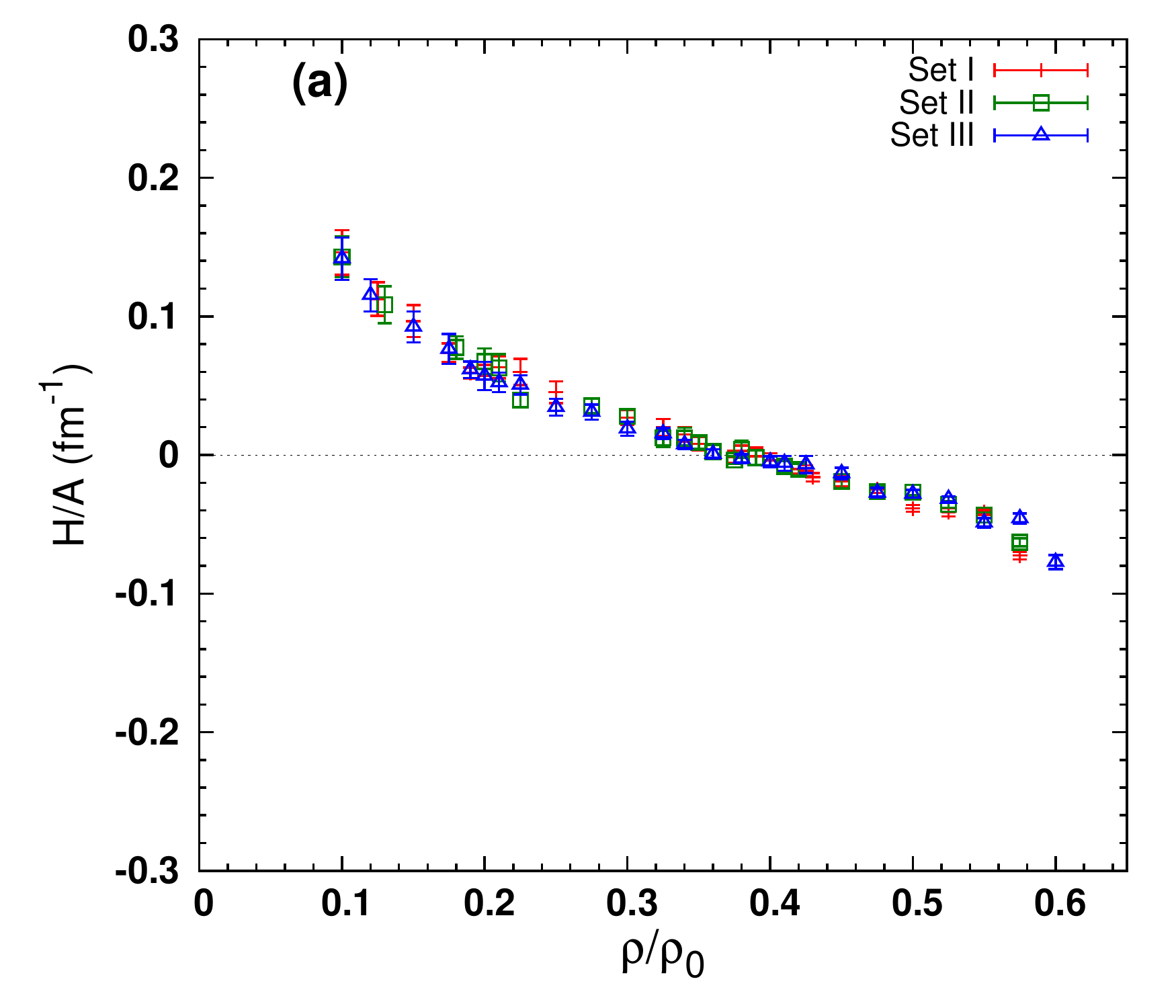}}&
   \resizebox{70mm}{!}{\includegraphics{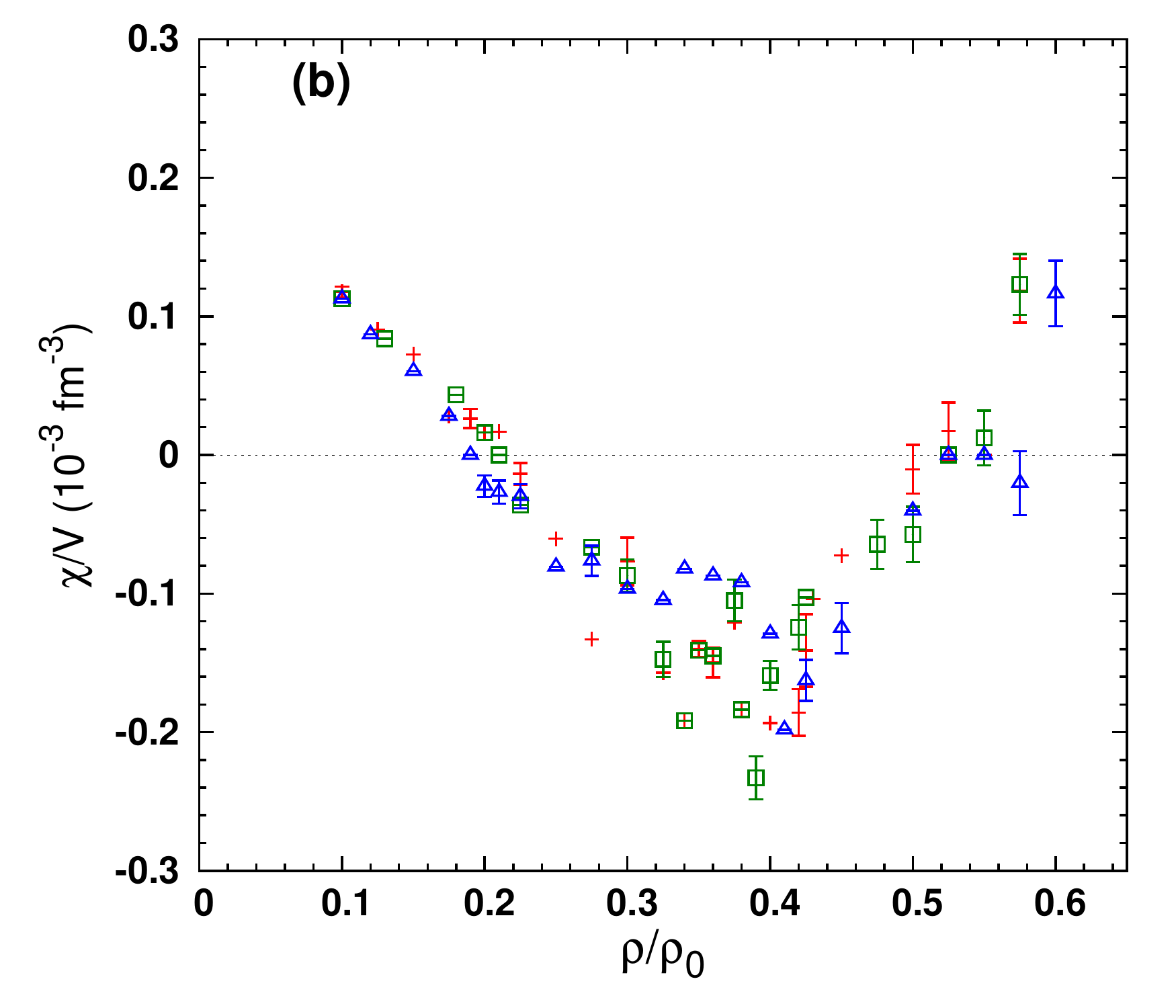}}
  \end{tabular}
  \caption{Comparison of normalised (a) mean curvatures and (b) Euler characteristics as functions of density for different $L$.}
  \label{f:mf}
 \end{center}
\end{figure}

Next, we compare the Minkowski functionals for three different values of the slope parameter $L$ corresponding to the parameter sets 
of Table \ref{symparams}, at $Y_p=0.3$. In Fig. \ref{f:mf} we plot normalised mean curvatures as well as the normalised Euler
characteristics as functions of normalised density for  all three parameter sets.
Error bars indicate the standard deviation in the range of $\rho_{\rm th}$, where $\chi/V$ has a plateau (see Fig. \ref{f:mf_th}).
From the figure we infer that although there are differences in the detailed behaviour of the Minkowski functionals (especially 
the Euler characteristics), overall they do not depend much on the parameter sets even if we take the estimated errors into
consideration. In terms of $L$ this means that the Minkowski
functionals are almost independent of its value.
But, this is in contrast to earlier calculations \cite{Bao14,Oyamatsu07,Grill12,Bao15,Sonoda08},
where it was found that the lower boundary of the pasta phase gets shifted to higher densities whereas the upper boundary is shifted to 
lower densities with increasing $L$.
The lower boundary is determined by the fission-like instability that increases with increasing volume fraction of the nuclear
region \cite{Oyamatsu07}. The volume fraction depends on $L$ through two competing factors. The first important factor is the 
saturation density, which for the asymmetric nuclear matter (within the parabolic approximation) is given by \cite{Oyamatsu07}
\begin{equation}
 \frac{\rho_s}{\rho_0} = 1 -\frac{3L}{K_0}\delta^2\, ,\label{e:rhos}
\end{equation}
where $K_0$ (280\,MeV in our case) is the incompressibility of symmetric nuclear matter. Eq. (\ref{e:rhos}) implies that for asymmetric 
matter at sub-saturation densities, the average density inside the nuclear region decreases with increasing $L$,
in turn leading to an increased volume fraction. Another controlling factor is the number of dripped neutrons, 
which increases with increasing $L$ at sub-saturation densities and causes the volume fraction to decrease. If the
second factor dominates over the first one, the lower boundary of the pasta phase gets shifted to higher densities.
However, for our case the lower boundary lies in the range  $0.1-0.125\,\rho_0$ for all three 
parameter sets i.e. for all different values of $L$. The 
dependence of the lower boundary on $L$ found in Ref. \cite{Sonoda08} might arise because of the difference in
the number of dripped neutrons for the two different models they use.
Therefore their result might be caused by adopting different nuclear models in
studying the $L$ dependence.

\subsection{Transition to uniform matter}
The upper boundary, which indicates the transition from pasta phase to uniform
nuclear matter, is sensitive to the symmetry energy. The symmetry energy at sub-saturation densities decreases
with increasing $L$ (see Eq. (\ref{esym2}) and thereby helps the transition to uniform matter to happen at lower 
densities. 
To determine the transition density from the pasta phase (spherical bubbles) to uniform matter we calculate the two-point correlation
function $\xi_{NN}$ for nucleon density fluctuations defined as:
\begin{equation}
 \xi_{NN} = \left<\bigtriangleup_N({\bf x})\bigtriangleup_N({\bf x} + {\bf r})\right>
\end{equation}
where the average is taken over the position $\bf{x}$ and the direction of $\bf{r}$ and $\bigtriangleup_N({\bf x})$
denotes the fluctuation of the nucleon density field $\rho_N(\bf{x})$ defined as
\begin{equation}
 \bigtriangleup_N = \frac{\rho_N({\bf x}) - \rho_{\rm av}}{\rho_{\rm av}}
\end{equation}
where $\rho_{\rm av} = {\cal N}/V$ is the average density of nucleons.
\begin{figure}
  \begin{center}
    \begin{tabular}{ccc}
      \resizebox{52mm}{!}{\includegraphics{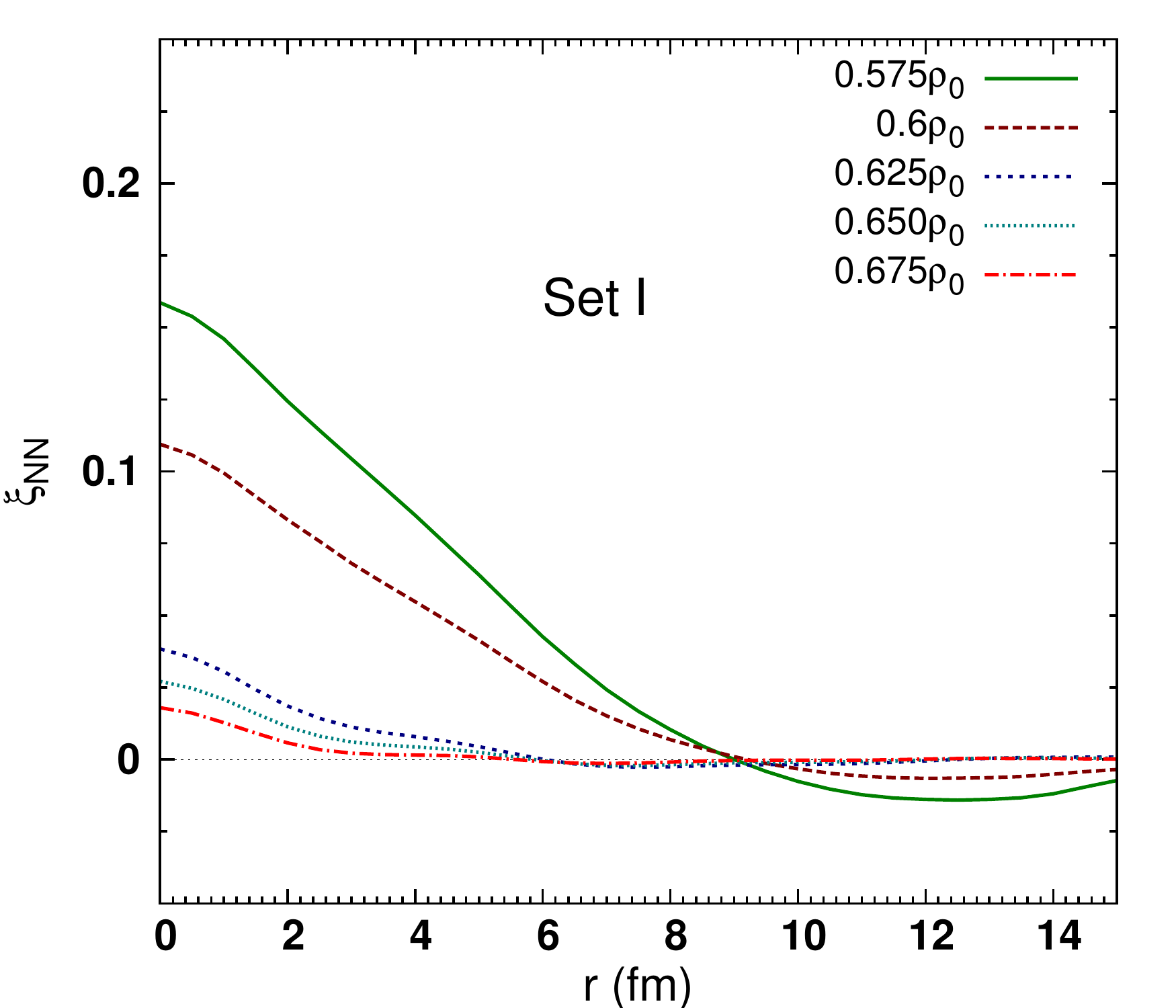}}&
      \resizebox{52mm}{!}{\includegraphics{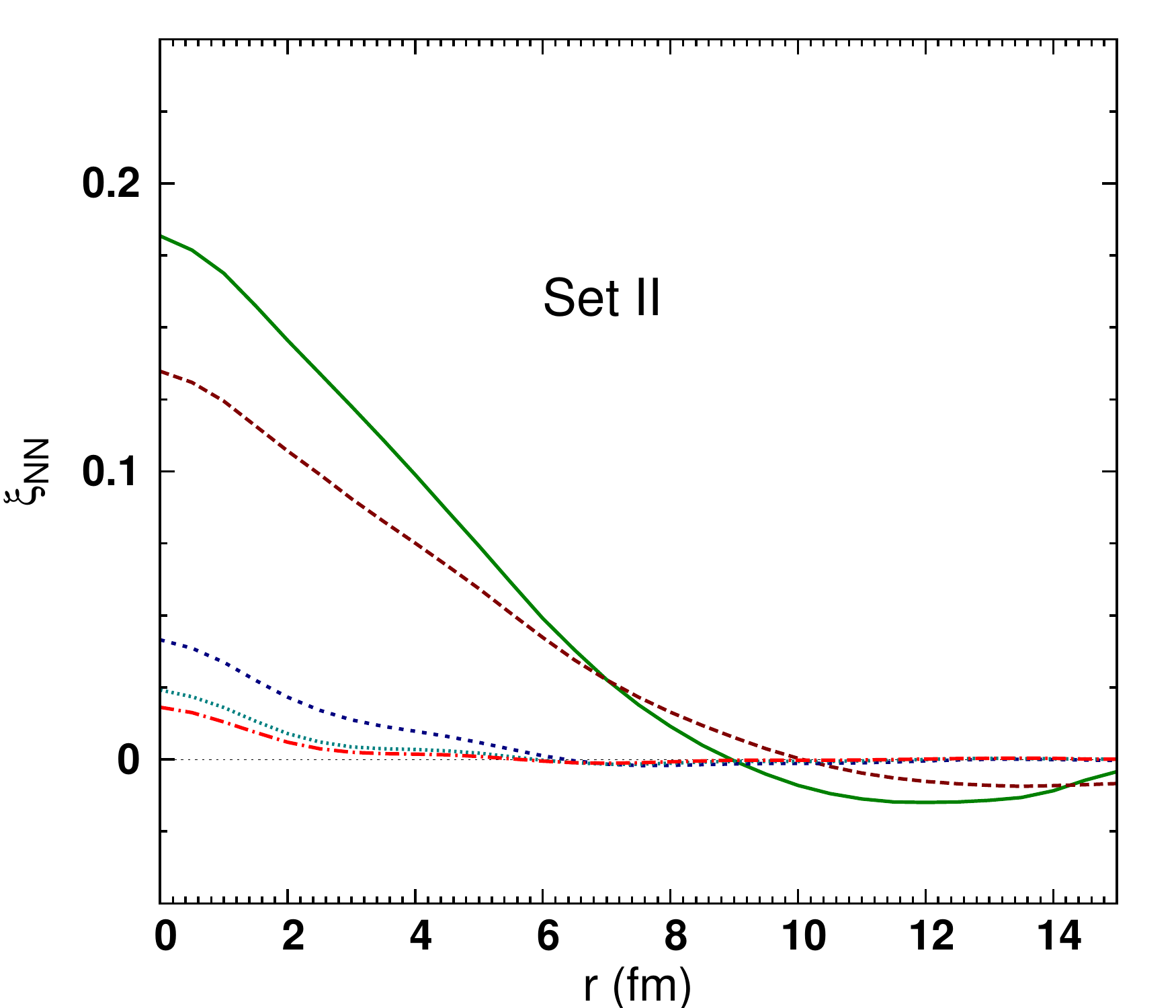}}&
      \resizebox{52mm}{!}{\includegraphics{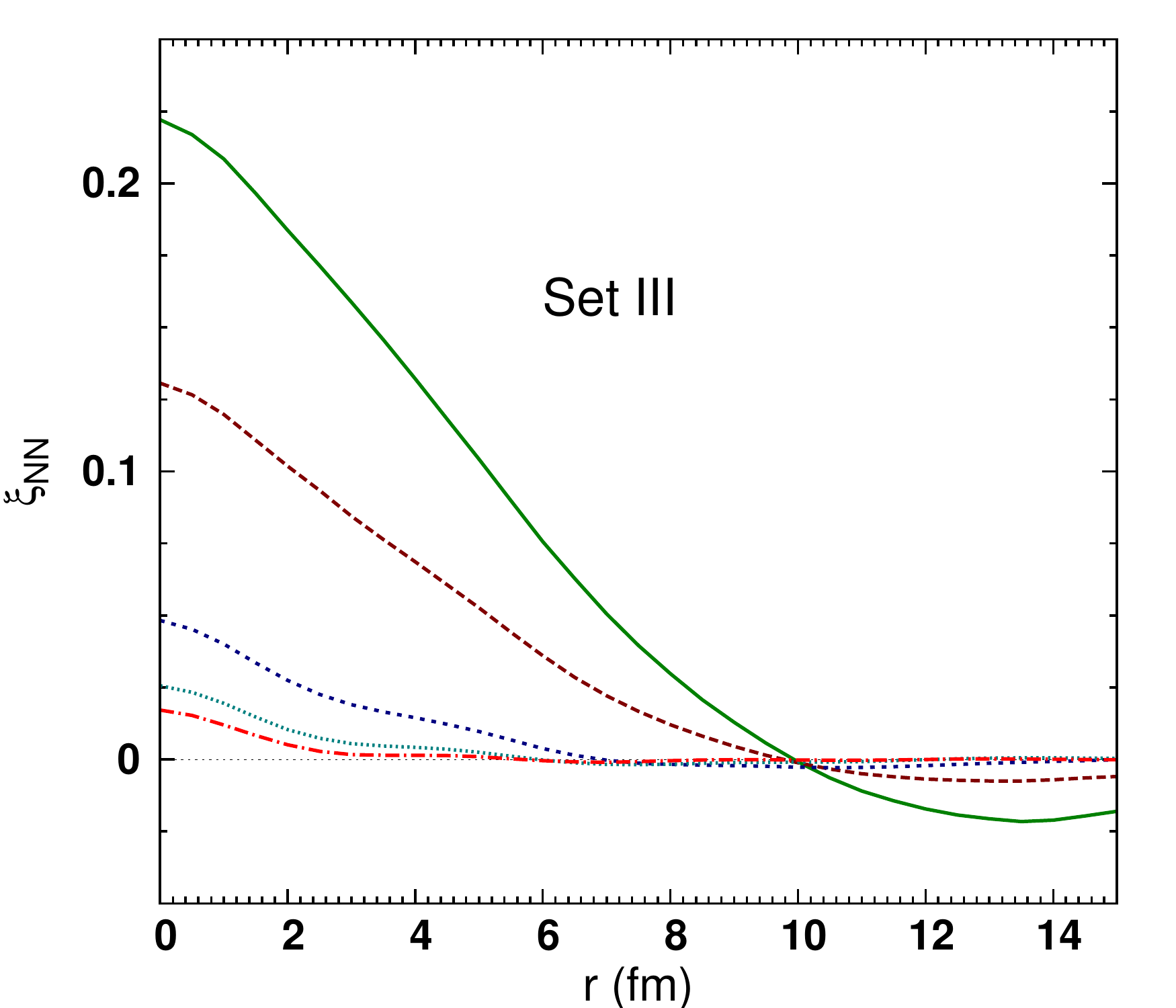}} 
    \end{tabular}
    \caption{Two-point correlation function $\xi_{NN}$ of the density fluctuation of nucleons at densities around the transition
             region of pasta to uniform matter.}
     \label{f:correl}
  \end{center}
\end{figure}
In Fig \ref{f:correl}, we plot the correlation function $\xi_{NN}$ in the density range
$0.575-0.675\,\rho_0$ . It can be observed that for all three cases long-range correlations vanish between $0.6\,\rho_0$ and
$0.625\,\rho_0$, indicating the transition from pasta to uniform nuclear matter. Moreover, the sudden vanishing of long-range correlations
points to the fact that the transition is of first order in nature for $Y_p = 0.3$.

\subsection{Improving on parabolic approximation}
Within the parabolic approximation, $e_{\rm asym}(\rho)$ in Eq. (\ref{asym}) simply equals to $e_{\rm sym}(\rho)\delta^2$. 
As the value of $e_{\rm sym}(\rho_0)$ is almost the same (see Table \ref{symparams}) for all three parameter sets, $e_{\rm asym}(\rho_0)$ also should be
equal for all cases for a given $\delta$. 
But Fig. \ref{f:asym} shows that $e_{\rm asym}(\rho_0)$ is quite different for different sets of parameter for $\delta\simeq0.41$. 
The figure also suggests the parameter set I to have the highest slope and the set III to have the lowest, at $\rho_0$. This is just the opposite 
of the calculated values of $L$ from Eq. (\ref{slope_sim}). This is the
result of the parabolic approximation we used for the determination of $e_{\rm sym}(\rho_0)$ and $L$ for nuclear matter with not such a small asymmetry. 
It was found from a systematic analytical study of the isospin dependence of the saturation properties of asymmetric nuclear matter
that the parabolic approximation is good for $\delta^2\leq0.1$ \cite{Chen09}.
Furthermore, from Fig. \ref{f:fit1} one can observe that the fits to the energy per nucleon are not satisfactory. 
All these observations lead us to  include an additional term in the expansion of the energy per nucleon as
\begin{equation}
 e(\rho,\delta) = e_0(\rho) + e_{\rm sym}(\rho)\delta^2 + e_{{\rm sym},4}(\rho)\delta^4 ,
 \label{epar4}
\end{equation}
where $e_{{\rm sym},4}$ is the fourth-order nuclear symmetry energy \cite{Chen09}. Expanding it around normal nuclear density to lowest order we get
\begin{equation}
 e_{{\rm sym},4}(\rho) = e_{{\rm sym},4}(\rho_0) + L_{{\rm sym},4}\chi \,
 \label{esym4}
\end{equation}
where $L_{{\rm sym},4}$ is the slope parameter of the fourth-order nuclear symmetry energy at $\rho_0$ given by
\begin{equation}
 L_{{\rm sym},4}=3\rho_0\frac{\partial e_{{\rm sym},4}(\rho)}{\partial \rho}\Bigg|_{\rho=\rho_0}\label{L4}.
\end{equation}

\begin{figure}
  \begin{center}
    \begin{tabular}{ccc}
      \resizebox{52mm}{!}{\includegraphics{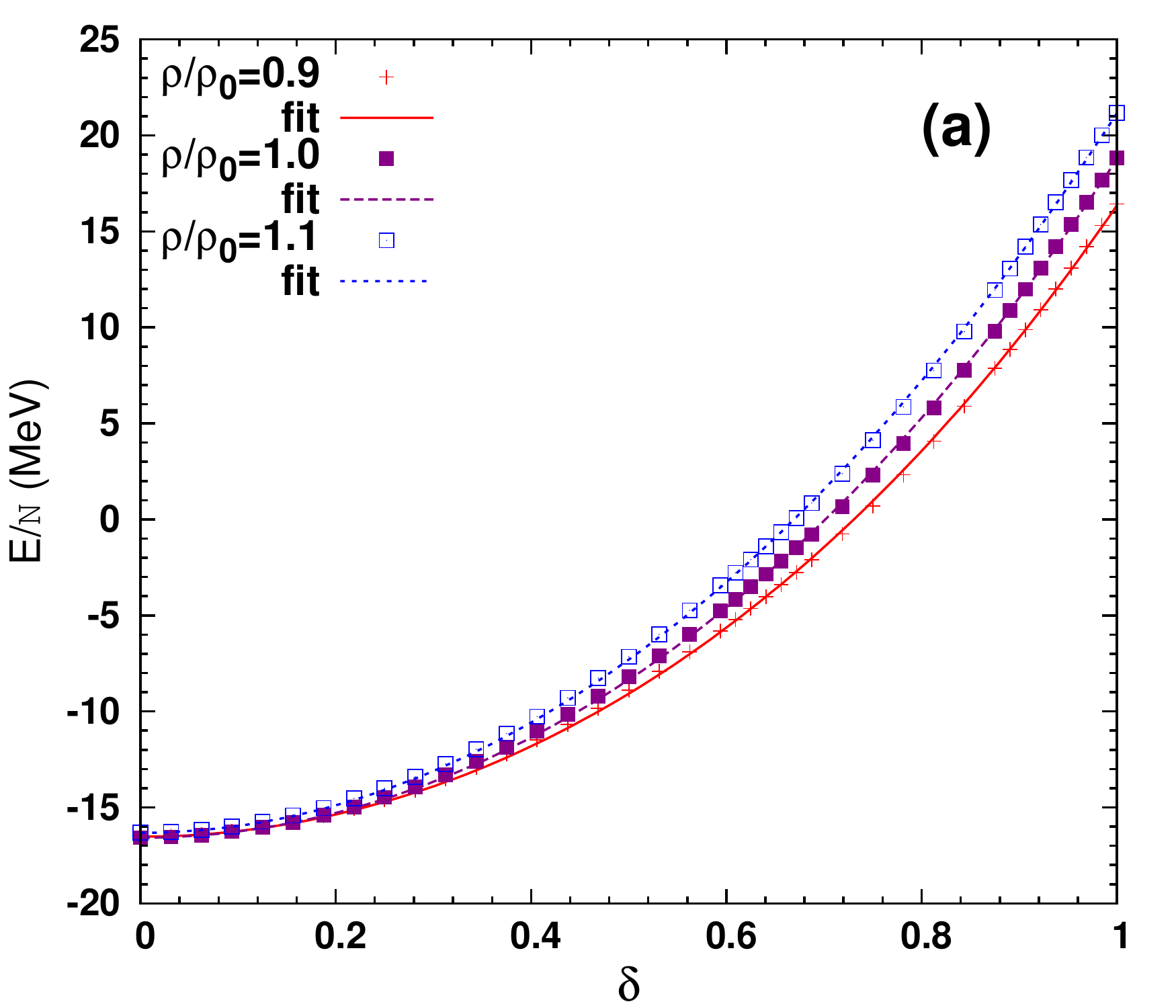}}&
      \resizebox{52mm}{!}{\includegraphics{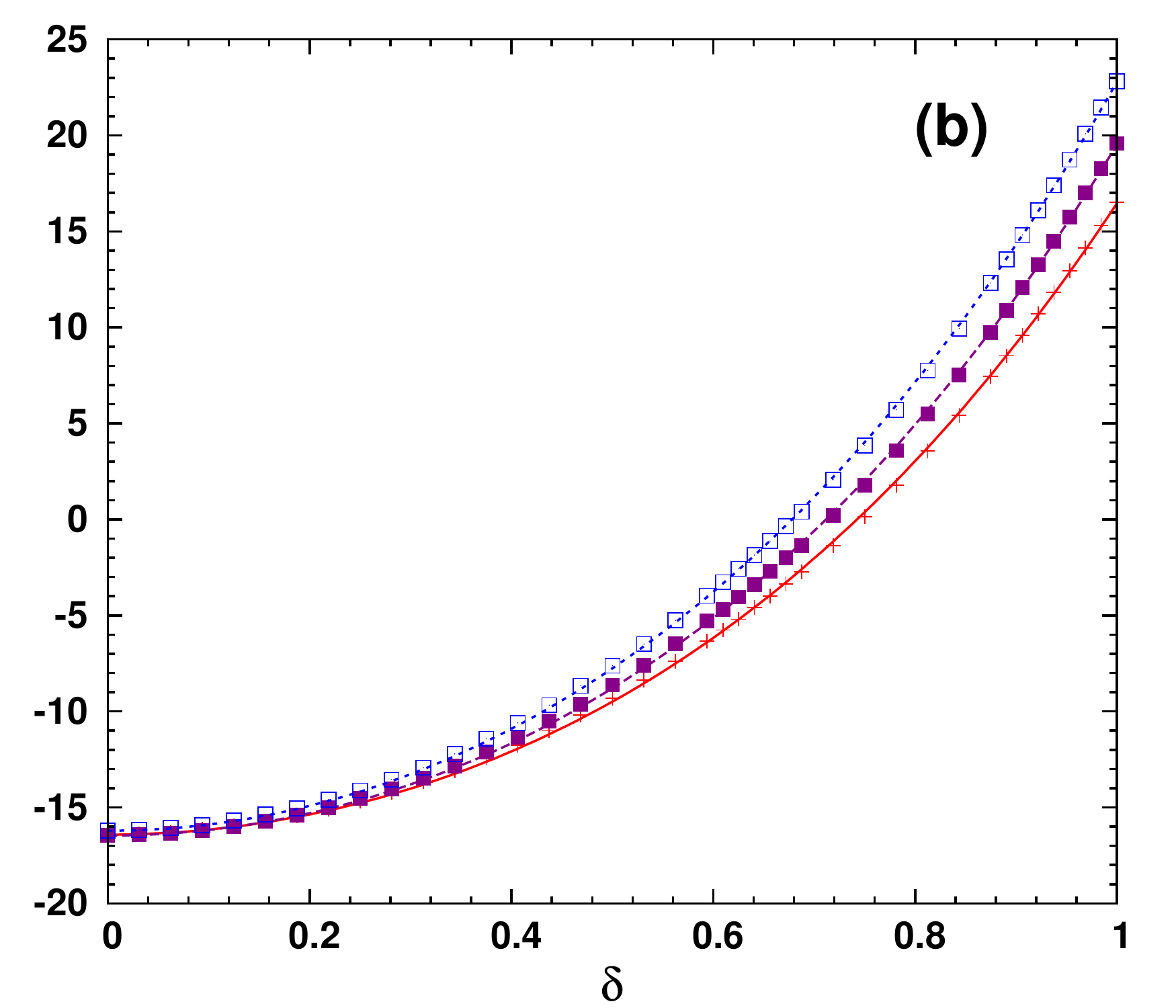}}&
      \resizebox{52mm}{!}{\includegraphics{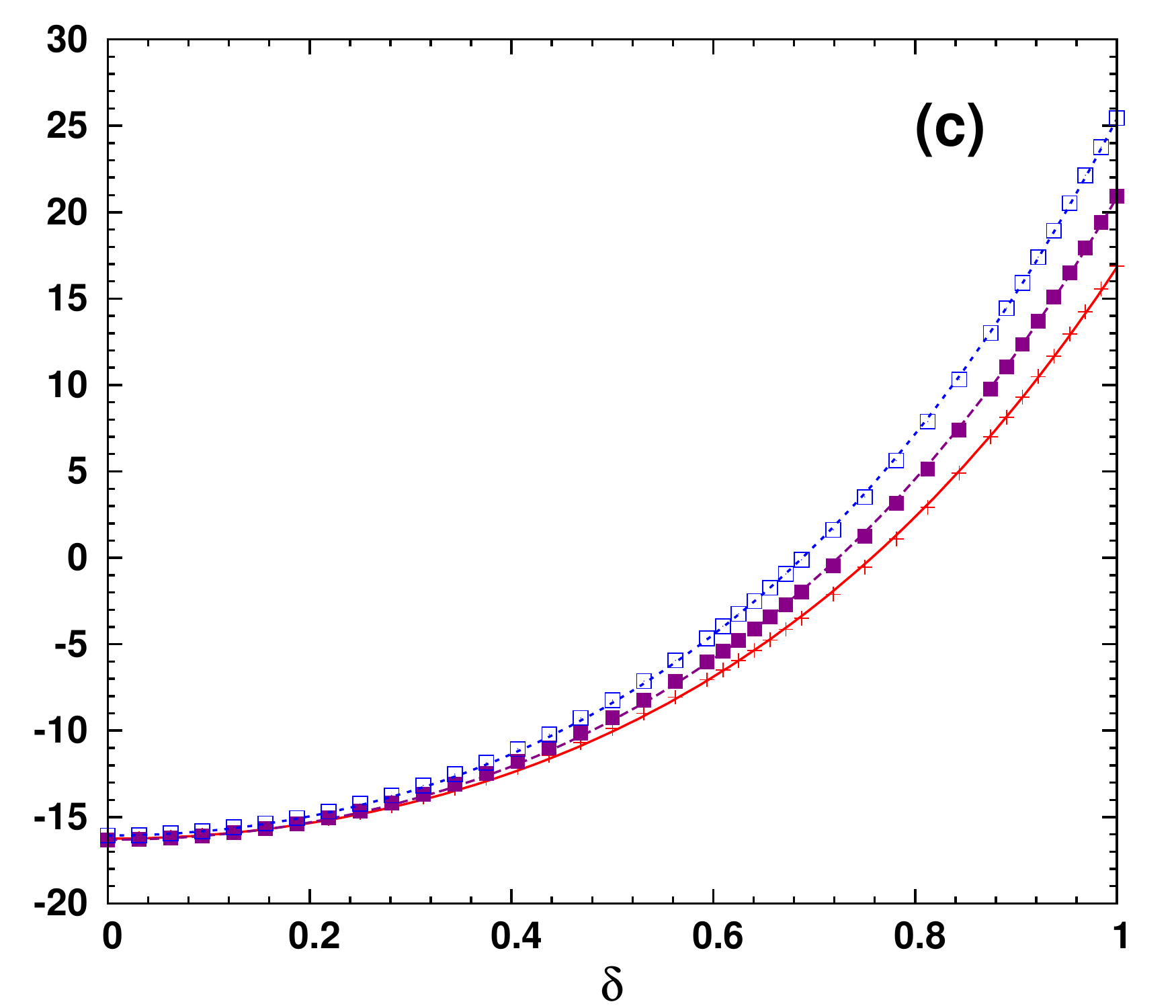}} 
    \end{tabular}
    \caption{Fit of the energy per nucleon using Eq. (\ref{epar4}) for different parameter sets}
     \label{f:fit2}
  \end{center}
\end{figure}
\begin{figure}
  \centering
   \includegraphics[width=0.6\textwidth]{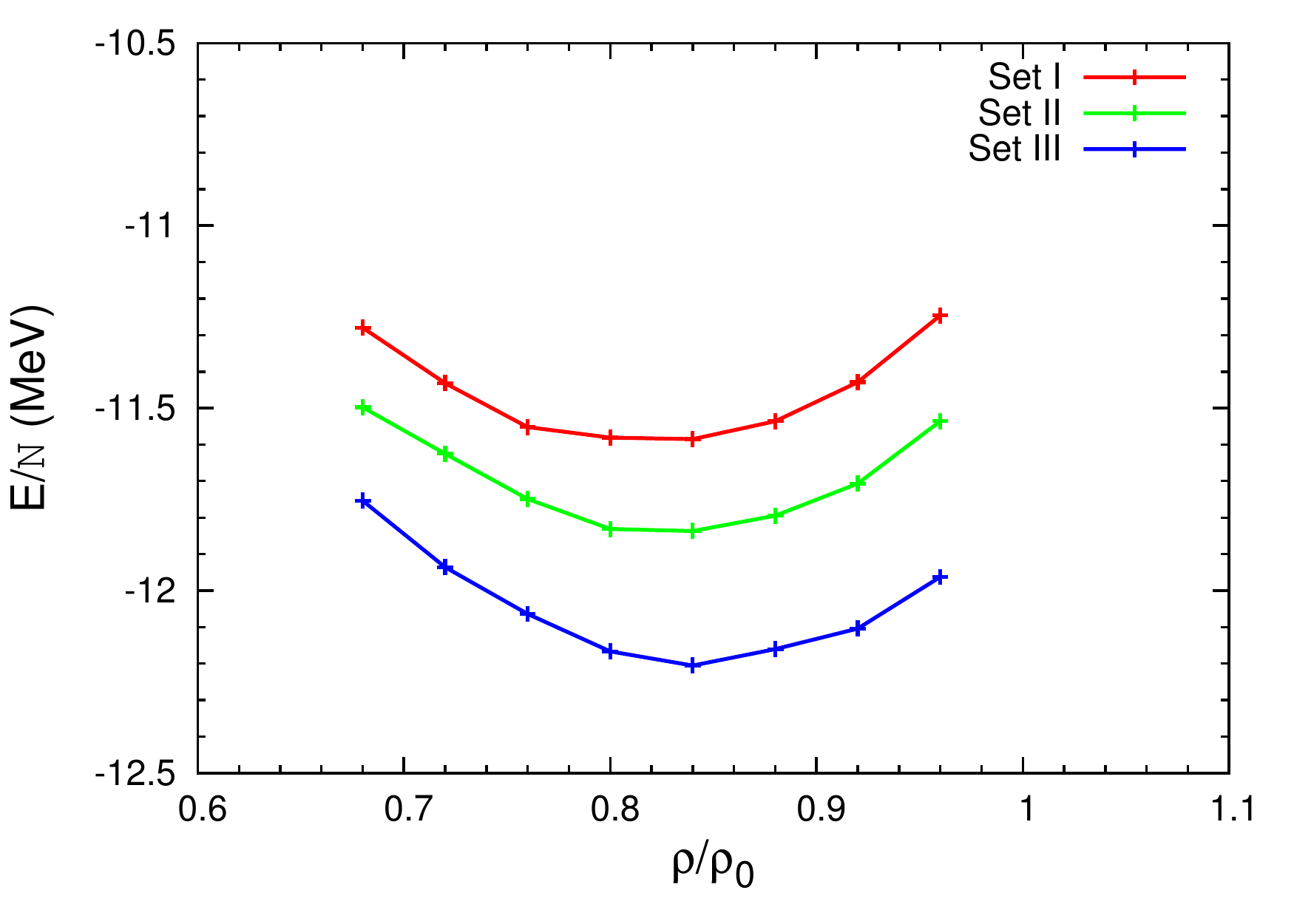} 
   \caption{Energy per particle as function of normalized density for $Y_p=0.3$ without Coulomb interaction. }
   \label{f:rhosat}
\end{figure}

We repeat fitting the data as described earlier but now with Eq. (\ref{epar4}). Fig. \ref{f:fit2} shows the 
obtained fits for all three sets. It is evident from the figure that the extended fits are much better, underlining 
the importance of a careful analysis of the numerical data. The resultant fit parameters are shown in
Table \ref{symparams4}.
\begin{table}[]
\caption{Symmetry energy coefficients}
\begin{tabular}{lccccc}
   & Set\qquad\qquad & $e_{\rm sym}(\rho_0)$(MeV)\qquad\qquad &$e_{{\rm sym},4}(\rho_0)$(MeV)\qquad\qquad & $L$(MeV)\qquad\qquad& $L_{{\rm sym},4}$(MeV) \\ \hline\hline 
   &  I   & $32.1$ & $3.27$ &$102.2$& $-33.2$\\
   &  II  & $28.9$ & $7.07$ &$91.7$& $0.0$\\
   &  III & $24.5$ & $12.7$ &$76.1$ & $50.0$\\   
 \end{tabular}
 \label{symparams4}
\end{table}
Similar fits are obtained for two other densities $1.1\rho_0$ and $0.9\rho_0$. Then $L$ is calculated using Eq. (\ref{slope_sim}) and $L_{{\rm sym},4}$
as below
\begin{equation}
  L_{{\rm sym},4} =  3\rho_0\frac{e_{{\rm sym},4}(1.1\rho_0)-e_{{\rm sym},4}(0.9\rho_0)}{1.1\rho_0-0.9\rho_0}\, . \label{slope4_sim}
\end{equation}
The resulting values are given in Table \ref{symparams4}. When we compare Table \ref{symparams4} with
Table \ref{symparams}, we see that the value of the symmetry energy $e_{\rm sym}(\rho_0)$ is no longer fixed 
around $\sim 34$ MeV, but varies in the range $\sim 24.5-32.1$ MeV. More interestingly, the values of
$L$ are now in opposite order for the same choices of parameters $C_s^{(1)}$ and $C_s^{(2)}$. All these values are now also
consistent with Fig .\ref{f:asym}. To establish it further we next use
these values of $L$ and $L_{{\rm sym},4}$ to calculate the saturation densities as \cite{Chen09}
\begin{equation}
 \frac{\rho_s}{\rho_0} = 1- \frac{3L}{K_0}\delta^2 -\frac{3L_{{\rm sym},4}}{K_0}\delta^4\, , \label{e:rhos4} 
\end{equation}
For $\delta\simeq0.41$ we obtain $0.828\rho_0$, $0.832\rho_0$ and $0.861\rho_0$ for Set I, II and III, respectively.
These values of saturation densities are very close to the values obtained from our simulation as shown in Fig. \ref{f:rhosat}.

\section{Summary and Conclusions}
We studied the inner crust of neutron stars within a quantum molecular dynamics approach.
Here, in particular we investigated the formation of pasta phases at densities close to the transition to
homogeneous matter. The interaction Hamiltonian was based on earlier work by \cite{Maruyama98}, where we extended
the isospin-dependent interactions to include non-linear terms in accordance with the isospin symmetric terms.
This allowed for a tuning of isospin-related features like the symmetry energy coefficient $e_{\rm sym}(\rho_0)$ and slope parameter $L$
while staying within the same model approach. We prepared three sets of parameters for the isospin-dependent interaction and obtained
very different symmetry energy behaviour. To check the reliability of these parameter sets we calculated the 
binding energies for the ground state of several nuclei and obtained reasonable agreement with the experimental values for all
of them. We derived $e_{\rm sym}(\rho_0)$ and $L$ for all parameter sets by fitting the numerical data to the expression of energy per nucleon 
written as Taylor series in neutron excess keeping both the lowest-order term as well as the next higher-order term. The lowest-order
approximation, also termed parabolic approximation, led to similar values of $e_{\rm sym}(\rho_0)$ but different values of $L$ for 
different parameter sets. On the other hand the higher order approximation produced different values for both $e_{\rm sym}(\rho_0)$ and $L$ 
for different sets. Careful investigation of all the simulation data revealed that the higher order term is necessary to have a 
correct description of the asymmetric nuclear matter with proton fraction typical for supernova environments.

To determine the dependence of the pasta phase on symmetry energy properties we studied the various pasta phases by determining 
the Minkowski functionals of the simulated nucleon distributions for all parameter sets representing different isospin forces. 
In contrast to previous molecular dynamics results \cite{Sonoda08}, but in agreement with static model
calculations \cite{Oyamatsu07}, the low-density onset of the pasta phase is quite insensitive to 
changing isospin asymmetry properties of the matter. The same holds
for the transition density from pasta phase to homogeneous
matter where we also have not observed any significant dependence for different isospin behaviour, unlike previous calculations
\cite{Bao14, Oyamatsu07, Grill12, Bao15, Sonoda08}. In conclusion, both the low density onset of the pasta and the transition density
to uniform matter are not sensitive to $e_{\rm symm}(\rho_0)$ and $L$. Furthermore, analysing two-point correlation functions we demonstrated
that the transition from the pasta phase to the core is fast, indicating 
a first-order transition for a proton abundance $Y_p = 0.3$. To see if these conclusions are affected 
by the finite size effects, as discussed in Ref. \cite{Molinelli15}, we plan to use larger system in future.

The numerical implementation of the simulation was done by making use of GPUs for the most time consuming parts 
of the calculation. With the computational framework in place, we will expand our simulations of the crust to 
study its transport properties and extend the simulations to substantially larger systems.

The authors are grateful to the referee for valuable suggestions which greatly improved the article. R. N. acknowledges financial 
support from the HIC for FAIR project and the NAVI program.

\end{document}